\documentclass[12pt]{article}
\usepackage{graphicx}
\begin{document}
\centerline{\large Asymmetries, Correlations and Fat Tails in Percolation Market Model}

\bigskip
\centerline {\bf Iksoo Chang$^1$, Dietrich Stauffer$^{1,2,3}$, and
Ras B. Pandey$^2$}

\bigskip
\noindent
$^1$ Department of Physics, Pusan National University, Busan 609-735, South
Korea

\noindent
$^2$ Department of Physics and Astronomy, University of Southern Mississippi,
Hattiesburg MS39406-5046, USA

\noindent
$^3$ Institute for Theoretical Physics, Cologne University, D-50923 K\"oln, 
Euroland
\vskip 1 cm
Abstract: Modifications of the Cont-Bouchaud percolation model for price
fluctuations give an asymmetry for time-reversal, an asymmetry between high
and low prices, volatility clustering, effective multifractality, correlations 
between volatility and
traded volume, and a power law tail with exponent near 3 for the cumulative
distribution function of price changes. Combining them together still gives
the same power law. Using Ising-correlated percolation does not change these
results. Different modifications give log-periodic oscillations before a crash,
arising from nonlinear feedback between random fluctuations.

\vskip 1 cm
\section{Introduction}

Fluctuations for stockprices or other markets have statistical properties
\cite{cont} which should be recovered by suitable microscopic models \cite{LLS}:
(i) There are only weak correlations between the price changes on successive
trading days; (ii) there are strong correlations between the absolute values of 
the price changes on successive trading days (``volatility clustering''); 
(iii)  the probability distribution function for the price changes (the
return histogram) decays with a power law in the tails \cite{lux,gopi}, with 
exponent near 3; (iv) price fluctuations show sharp peaks and flat valleys,
i.e. a high-low asymmetry \cite{roehner}; (v) price fluctuations are not
invariant against time reversal, i.e. they show a forward-backward asymmetry
\cite{muzy}; (vi) for long times (many weeks) a crossover to a more Gaussian
return histogram is seen \cite{gopi}; (vii) the $q$-th moments of the return 
histograms show multifractality, i.e. their time exponents are not a linear 
function of $q$.

The Cont-Bouchaud model \cite{cb} uses random percolation clusters as groups
of traders buying or selling together and thus simulates human herding.
In the simple version, at each iteration each cluster buys with probability
$a$, sells with probability $a$, or sleeps with probability $1-2a$. This
activity thus measures the time between two iterations: small $a$ mean 
short times and large $a$ mean long times. Price changes (more precisely,
changes in the logarithm of the price, where the prices are measured in units
of some fundamental price) are proportional to the difference between
supply and demand. By construction, on average one has as many buyers as 
sellers, and prices go up or down with equal probability and without 
correlations between consecutive steps, fulfilling (i). Increasing $a$ from low
values to its maximum value 1/2 changes the return histogram from power-law
tails to a more Gaussian shape \cite{penna}, fulfilling property (vi). Changing
$a$ proportionally to the last price change fulfills properties (ii) and (iv).
Making the ratio of buying and selling probabilities different from unity
and depending on the last known price and price change \cite{chang2} fulfilled 
(v) and even gave log-periodic oscillations \cite{sojo} after the crash of a 
price bubble
\cite{pandey}. Averaging over concentrations both near to and far away from the
percolation threshold together with a square-root dependence of price changes
on the difference between supply and demand gave a cumulative histogram of
returns $r$ (= relative price change) with tail decaying asymptotically as
$|r|^{-\mu}$ where $\mu = 2(\tau+\sigma-1) \simeq 3$, fulfilling (iii).  
Thus all desired properties (i) to (vi) were fulfilled by one or the other
modification; we are not aware of another microscopic model fulfilling all.
(Perhaps the Levy-Solomon-Huang model can also give all these properties
\cite{huang}.)

The aim of the present work is to combine these modifications and to check
which properties are then still valid in this combination model. We also
check the influence of Ising correlations on the traders (occupied sites)
on the lattice \cite{ising}.

\section{Model}
An $L \times L$ triangular lattice (known to be in the same universality class 
as the more often studied square lattice) is occupied randomly (or later 
with Ising correlations) with probability $p$, which in our simulations does
not exceed the percolation threshold $p_c=1/2$ at which an infinite cluster is
formed. Clusters are sets of neighbouring occupied sites; right at $p=p_c$ the 
number $n_s$ of large clusters containing $s$ sites decays as $1/s^\tau$ while 
near $p_c$  it is $s^{-\tau} f[(p-p_c)s^\sigma]$ according to standard
percolation theory \cite{books} with $\tau \simeq 2, \; \sigma \simeq 0.4$
in two dimensions. 

For market simulation, each occupied site is regarded as an agent (trader,
investor), and clusters are companies of traders buying and selling together
an amount proportional to the number $s$ of agents in the cluster. Each
cluster is active with probability $2a$ and inactive with probability $1-2a$
during each time step independently; if it is active it buys with probability
$p_b$ and sells with probability $p_s = 1-p_b$, where $p_b = p_s = 1/2$ only
in the simple unbiased version \cite{cb}. Thus each cluster buys, sells, or
sleeps with probabilities $2ap_b, \; 2ap_s, \; 1-2a$, respectively. The
total demand (supply) is $\sum_s n_ss$ with the sum running over all
buying (selling) clusters and excluding the largest cluster. The price change 
$r$ at one time step is proportional
to the difference of demand and supply, or to the square root \cite{zhang}
of that difference. We interpret the price change as a relative one, more 
precisely, this return $$r(t) = x(t)-x(t-1)$$ is proportional to the change in 
the logarithm $x$ of the price (in units of an initial or fundamental price) 
and thus together with $x$ fluctuates about zero, in arbitrary units. In the
simplest model we assume
$$ r = \sum_{buy} n_ss - \sum_{sell} n_ss \quad .$$
We take into account the agent psychology who believe that past trends will
continue, but also their more reasonable wish to buy (sell) if $x$
is low (high). Thus if $r < 0$ ($r$ and $x$ are measured at the previous time 
step) we take at first
$$ p_b = 0.5 - 5 \cdot 10^{-7} x + 5 \cdot 10^{-4} r \quad ,\eqno (1)$$
while for $r > 0$ we take 
$$ p_b = 0.5 - 5 \cdot 10^{-7} x + 5 \cdot 10^{-5} r \quad .\eqno (2)$$
The difference between these two equations is one order of magnitude for
their last term and takes into account that agents are risk adverse and thus
more impressed by a downturn than by an upturn of the market. With these
parameters at $p=p_c$, a reasonable slight asymmetry between up and down for
the prices was simulated \cite{chang2}. 

To avoid the assumption that agents act only at the percolation threshold, we
follow \cite{sornette} and sum up over all $p$ between zero and the threshold,
in units of one percent; and we take Zhang's square-root law \cite{zhang}:
$$ r^2 = \sum_{buy} n_ss - \sum_{sell} n_ss \eqno(3) $$
($|r|$ is then rounded downward to an integer).
With these assumptions but without the bias of eqs(1,2) (i.e. with $p_b=p_s = 
0.5$) the probability $P(r)$ for a change $r$ decays as 
$r^{1-2\tau-2\sigma} = 1/r^{3.9}$ as desired \cite{lux,gopi}, for $r \gg 1$
(and $r$ much smaller than an upper limit given by cluster radius $= L$).

\begin{figure}[hbt]
\begin{center}
\includegraphics[angle=-90,scale=0.5]{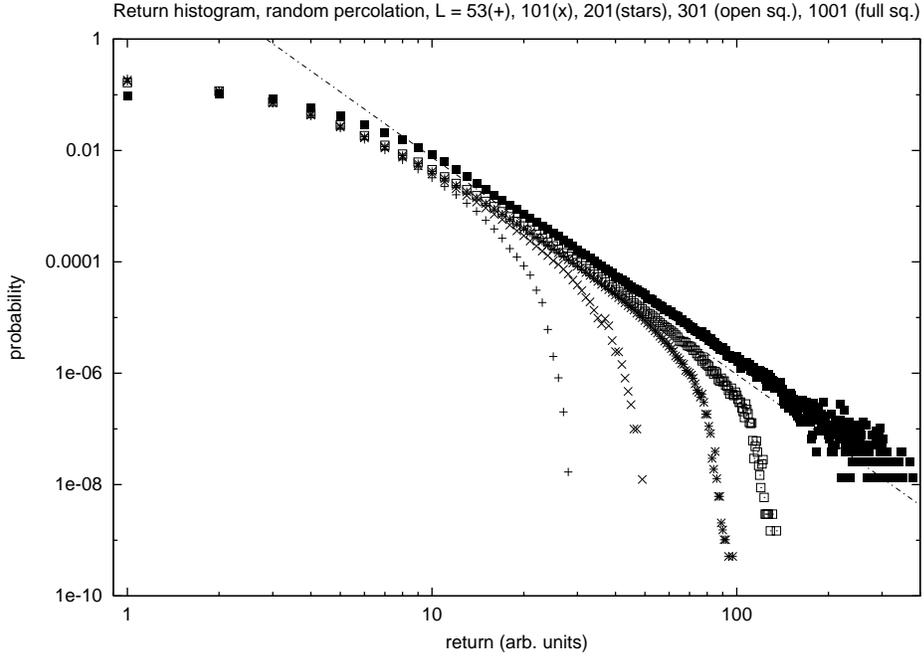}
\end{center}
\caption{
Probability $P(|r|)$ to have a return $|r|$ averaged over positive
and negative values, for various $L$ as given in the headline. We simulated
1000 lattices for $L= 53$ and 101, 20000 for 201, 6400 for 301, and 640 for 
1001. For each lattice, 5000 time steps were made where each cluster decided
to buy, sell or sleep.
}
\end{figure}    

\begin{figure}[hbt]
\begin{center}
\includegraphics[angle=-90,scale=0.5]{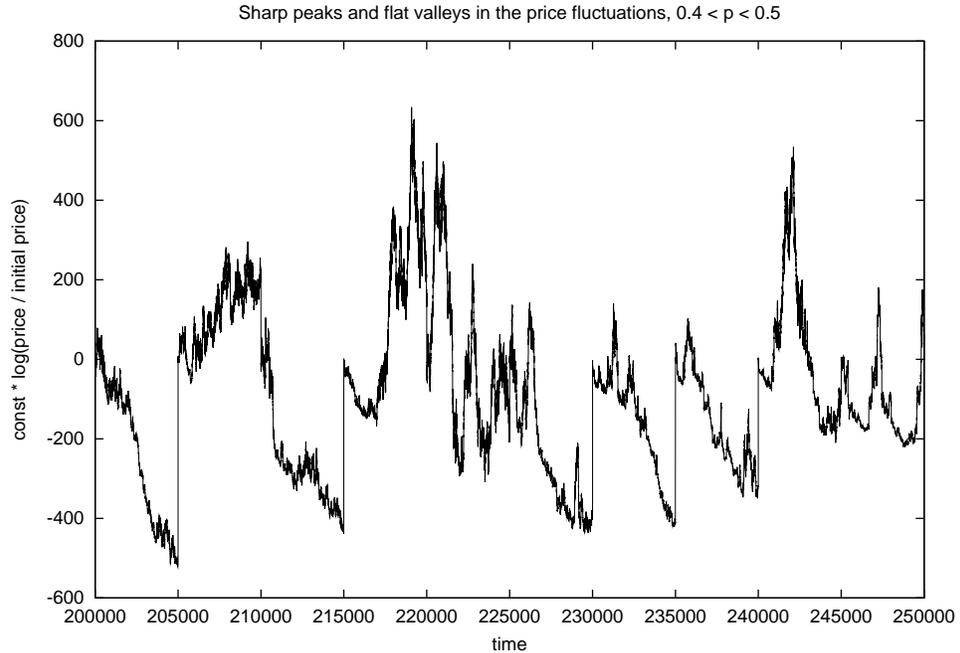}
\end{center}
\caption{
Last part of the price versus time curve for $L=301$. For each of
the 50 concentrations $p$ up to $p_c = 50$ percent, 5000 time steps were 
simulated for one lattice. The straight line has the theoretical slope 3.9.
}
\end{figure}    

\begin{figure}[hbt]
\begin{center}
\includegraphics[angle=-90,scale=0.5]{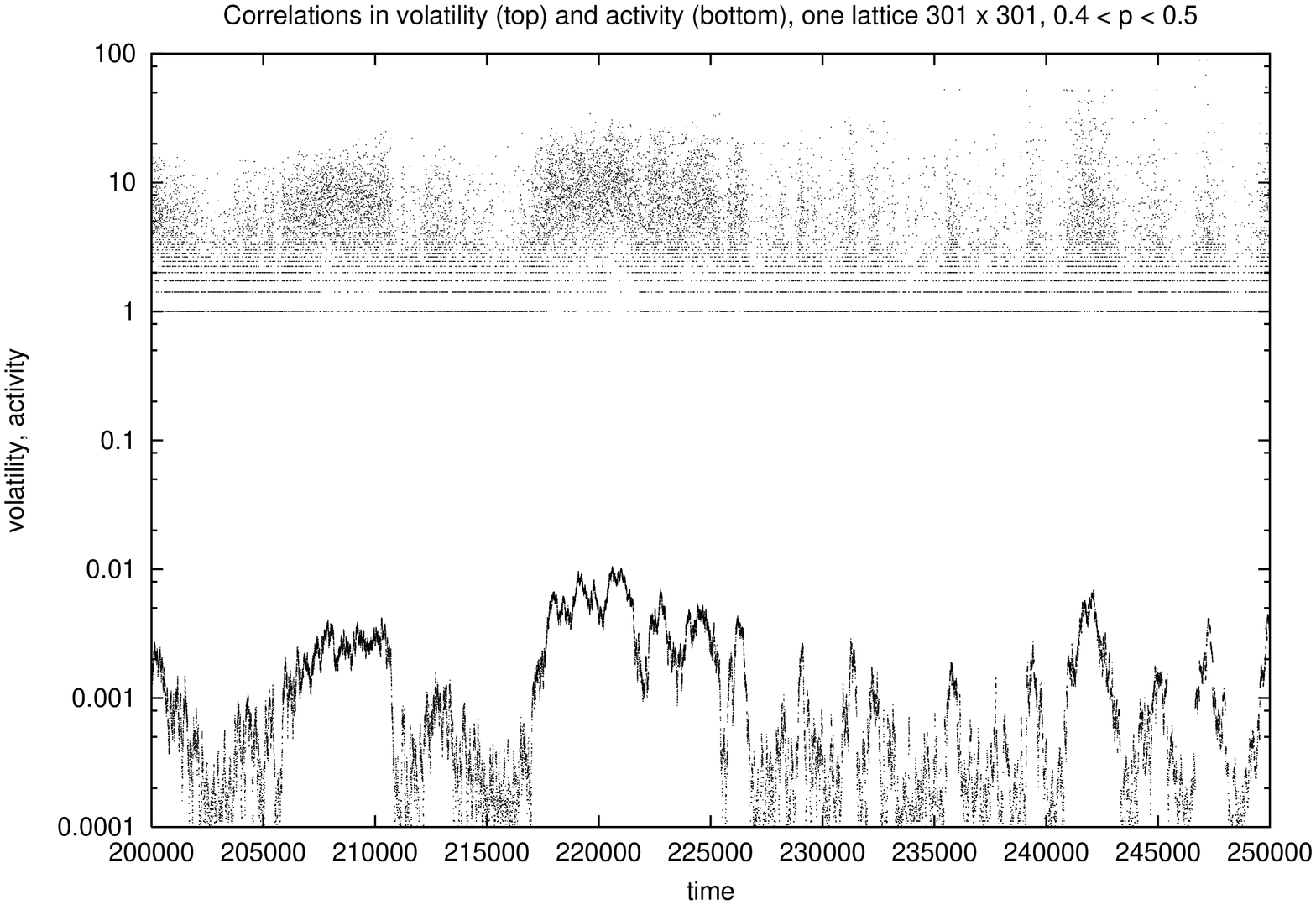}
\end{center}
\caption{
Same simulation as in Fig.2, but returns $r$ (top) and activity $a$ 
versus time. When the activity is large, the price changes are also large.
Moreover, times with large $a$ and $r$ cluster together. 
}
\end{figure}

Assuming \cite{jan}, in accordance with reality \cite{roehner2}, the activity 
$a$ to increase with positive returns $r$, 
$$ a(t+1) = a(t) + 0.5 \, r(t)/L^2 \eqno (4) $$ 
(but always $1/10^4 < a < 1/2$) gives sharp peaks compared to flat valleys in 
$x(t)$ \cite{roehner} and volatility clustering \cite{jan}, at $p=p_c$ without 
the complications of eqs(1-3). 

For Ising correlated agents \cite{ising}, we no longer distribute agents
randomly on a lattice but identify them with up spins of the standard 
Ising model at temperature $T$. Earlier\cite{ising}, this temperature was
varied between $T_c$ and infinity, leading always to $p = 1/2$ for the 
fraction of occupied lattice sites. The results were similar to random
occupation. Now, closer to \cite{sornette} we take
much smaller $p$ by increasing $T$ from $0.9 T_c$ to $1.01 T_c$ and thus 
$p$ from 0.06 to about 0.5, starting with all sites empty in Glauber kinetics.
(Since the spontaneous magnetization $1/2-p$ at $T=T_c$ is about $1/L^{1/8}$
and thus still quite large for realistic market sizes $L$, we took the
upper limit at $1.01 T_c$ to push $p$ closer to 1/2.)

Now we check if a combination of all these modification still gives reasonable 
results.

\section{Results}

We combine modifications (1-4) and sum up over random occupation probabilities 
from one to 50 percent. Fig.1 shows that for large enough markets the tail 
exponent is close to the desired 3.9. Fig.2 contrasts the resulting
sharp peaks with the flatter valleys of $x(t)$. Fig.3 shows volatility
clustering, i.e. large values of $|r|$ have a tendency to cause large $|r|$
thereafter. (If $r$ is not plotted logarithmically, the results are symmetric
about zero.) The same figure shows that large $r$ are correlated with 
large activity $a$ in agreement with reality \cite{gopi2}. Unfortunatly, the
asymmetry in time-reversal \cite{chang2} is destroyed by assumptions (3) or
(4); when we make the small prefactor of $r$ by which eqs(1,2) differ ten times 
larger, some asymmetry is seen again.

The probability of a downward movement followed by an upward movement
\cite{zhang} is diminished by one to two percent compared to the three
other choices, for 1000 lattices of $L = 53$ and 101.

The crossover to a more Gaussian $P(r)$ for increasing activity $a$ is also 
no longer seen since now a large $a$ only means a large {\it initial}
activity. Due to modification (4), $a$ decreases about linearly with 
time during the first half of the simulation (small $p$, small clusters and
thus small $r$), and then fluctuates strongly near $1/10^4 \dots 1/10^2$ when
market fluctuations become important. These low activities in the second half
destroy the Gaussian behaviour. 

For Ising-correlated agents the results are similar; Fig.4 shows the fat tails.
\begin{figure}[hbt]
\begin{center}
\includegraphics[angle=-90,scale=0.5]{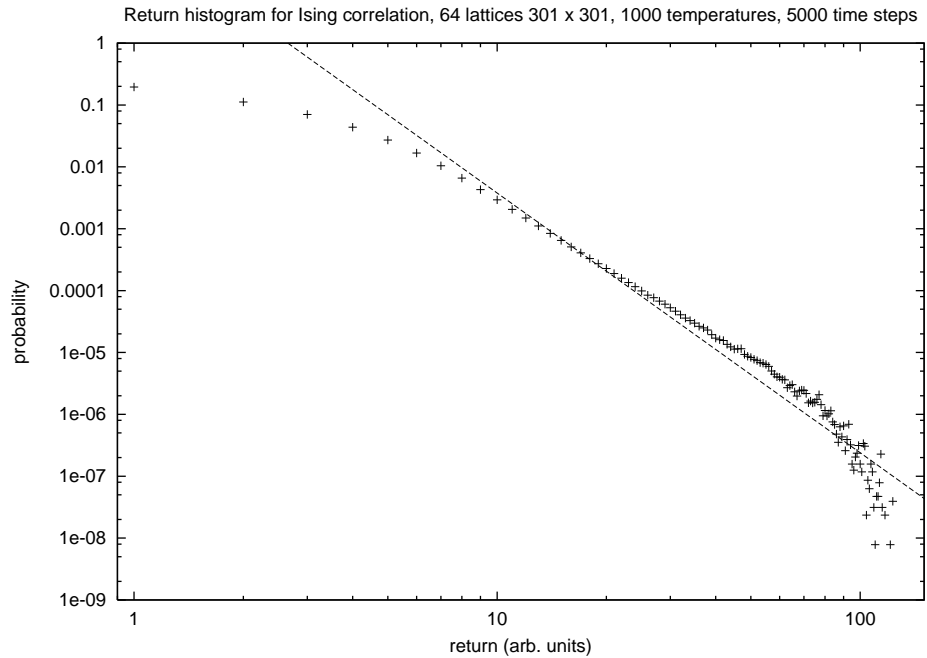}
\end{center}
\caption{
Similar to Fig.1, but for Ising correlation and 1000 temperatures
for $L=301$. The straight line has the theoretical slope 4.2.
}
\end{figure}    

\begin{figure}[hbt]
\begin{center}
\includegraphics[angle=-90,scale=0.5]{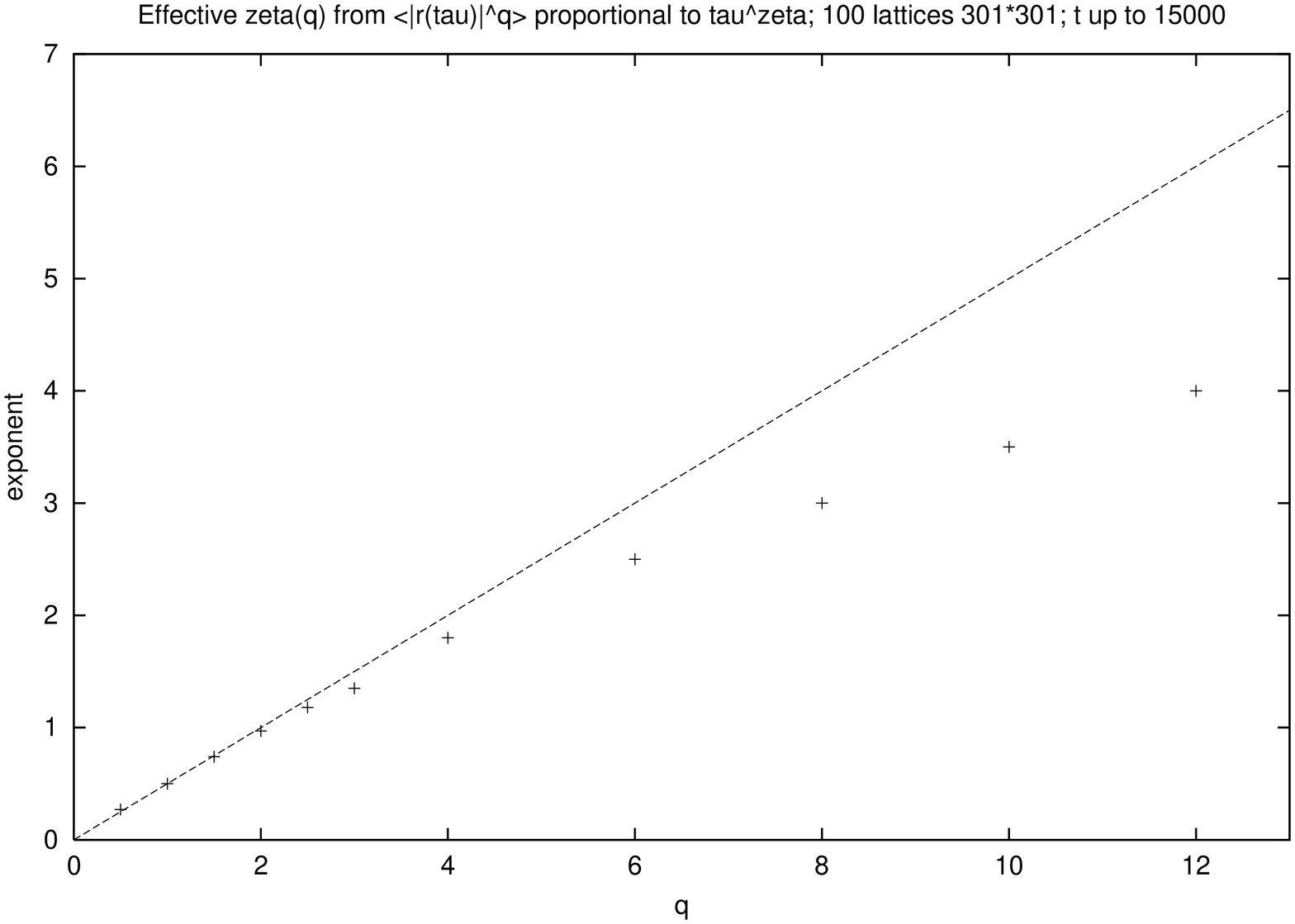}
\end{center}
\caption{
Effective exponents $\zeta$ versus $q$, from $10^3 < \tau < 3750$, for 
the moments $<|r(\tau)|^q> \propto \tau^\zeta$ with $r(\tau) = <x(t+\tau) - 
x(t)>_t$. The straight line $\zeta = q/2$ would hold for random walks, while
the downward curvature shows (effective) multifractality. 
}
\end{figure}    

Multifractality \cite{ausloos} here refers to the $q$-th moments
$<|x(t+\tau) - x(t)|^q>_t \propto \tau^\zeta$ where 
for each $p$ separately we average over long times
$t$. If $\zeta = \zeta(q)$ is linear in $q$ we have usual scaling; otherwise
we have multifractality (multiaffinity, multiscaling, property (vii)). Fig.5 
suggests multifractality in
even better agreement with reality than for the unmodified Cont-Bouchad model
\cite{casti}. In both cases the exponents are effective and might differ from 
the asymptotic behaviour \cite{bouchaud}.

In summary, properties (i) to (iv) are recovered, while the asymmetry (v) is 
partially and the crossover towards Gaussians is totally destroyed. 

\section{Log-periodic precursors of market crashes} 

Before or after a stock market crash, one may observe log-periodic oscillations
of the price. Major crashes are quite rare, perhaps several in a century for
one given market. They may be outliers \cite{joso} not described by the above 
model for normal behaviour. Thus we use now a related but different model for
log-periodicity.

One of the microscopic models to show them was our {\it non-linear}
restoring and inertia force \cite{pandey} : If prices go up (down), 
people have the tendency to buy (sell); this herding (inertia, hysteria,\dots)
enhances the trend and destabilizes the market. On the other
hand, when prices are high, people have a tendency to sell, while they prefer
to buy if prices are low; this restoring force stabilizes the market. When 
the restoring force was nonlinear, proportional to the fifth power of the
deviation from the perceived fundamental price, and when the market started
with a far too high price, then several oscillations were observed whose
period increased exponentially with the order of the oscillation, until random
noise took over:
$$ \log(y) = x \; \propto \; \cos(\lambda \log t)*D   \eqno (5)$$
where $y$ is the ratio of the actual price at time $t$ to the constant
fundamental price, and $D$ is a damping factor, like $\log D \propto -t$.
The actual price changes were produced by selling and buying clusters in
a Cont-Bouchaud type model, where instead of geometrical clusters simply
step sizes $s$ with a probability $n_s = N/s^{5/2}$ (Levy walks) were
assumed.

\begin{figure}[hbt]
\begin{center}
\includegraphics[angle=-90,scale=0.5]{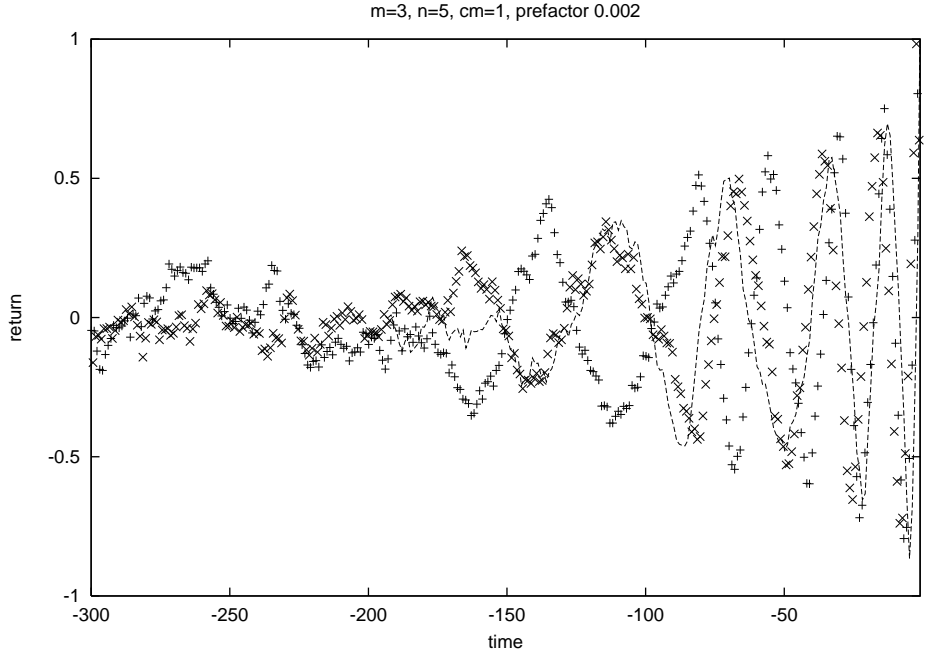}
\end{center}
\caption{
Three examples of random nonlinear behaviour, eq.(6) plus noise,
leading to a crash; only the random numbers differ.
}
\end{figure}    

\begin{figure}[hbt]
\begin{center}
\includegraphics[angle=-90,scale=0.5]{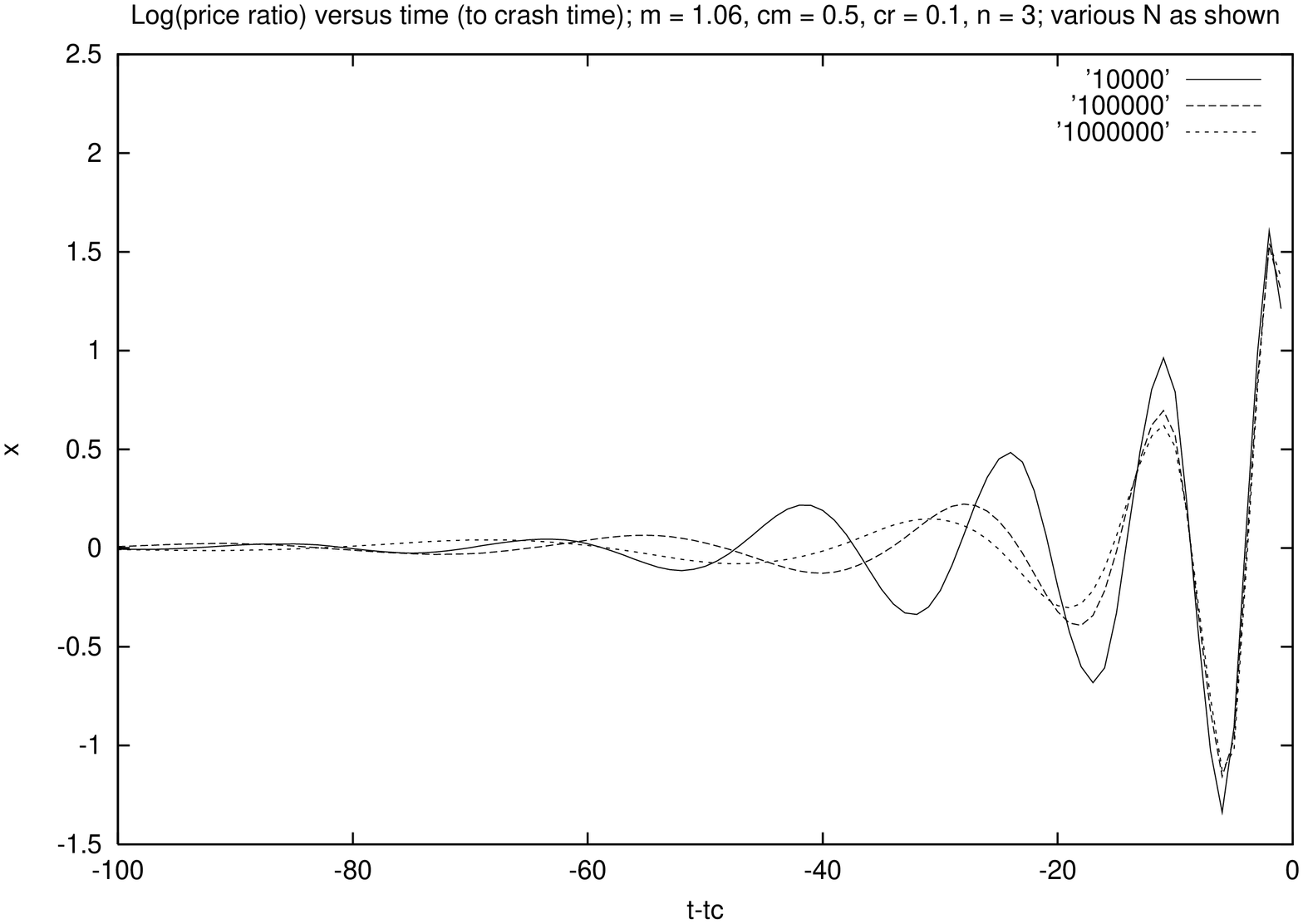}
\end{center}
\caption{
Average over thousand samples of the full model, for various market sizes
as shown.
}
\end{figure}    
A more phenomenological and simpler, but also more general approach \cite{ide}
used the differential equation
$$ d^2x/dt^2 = c_m (dx/dt)^m - c_n x^n \eqno (6) $$ 
with not only the restoring force $x^n$ but also the inertia $(dx/dt)^m$ being
nonlinear. (If $m$ and $n$ are not odd integers, suitable signs and absolute
values have to be used.) Now a small perturbation may grow until $x \rightarrow
\pm \infty$ at some finite crash time $t_c$. We added a random noise 
$\pm 0.002$ to the RHS of this differential equation and then found results
as in Fig.6, where three samples differing only in their random number
sequences for the noise are plotted versus $t-t_c$. We see that shortly before
the singularity they show roughly the same maxima and minima, with the time 
between the extrema decreasing when $t$ approaches $t_c$. For earlier times
the noise dominates, and the three curves differ significantly.

We apply these new ideas \cite{ide} to our old model \cite{pandey} 
and take as probability $p_b$ to buy:
$$ p_b = 0.5 - c_n x^n +  c_m (dx/dt)^m  \eqno (7) $$
where the derivative in this discrete model means the price change from one 
iteration to the next. Thus each cluster sleeps with probability $1-2a$,
and if it does not sleep it buys with probability $p_b$ and sells with
probability $1 - p_b$, where $0 < a < 1/2$ is the activity. Initially, $x$ and
$dx/dt$ are zero; prices and their changes are normalized by $N$.
We then see for suitable parameters, like $a = 1/4,\; 
m$ slightly above 1
and $c_m = 0.5$ ($c_n = 0.1$ to 1 and $m = 2$ to 5 seem less crucial), that
after $10^2$ to $10^3$ iterations of random fluctuations, some oscillations
emerge which become faster and faster, until we reach a crash at $t=t_c$ 
defined as $p_b=0$. Fig.7 shows average prices over 1000 samples, with 
market size $N = 10^4 \dots 10^6$, where the random fluctuations mostly
cancel each other while the oscillations shortly before the crash do not
cancel. (The crash times are log-normally distributed.) 
Fig.8 shows the resulting times where $x=0$; for large $N$ they indicate 
log-periodicity.

\begin{figure}[hbt]
\begin{center}
\includegraphics[angle=-90,scale=0.5]{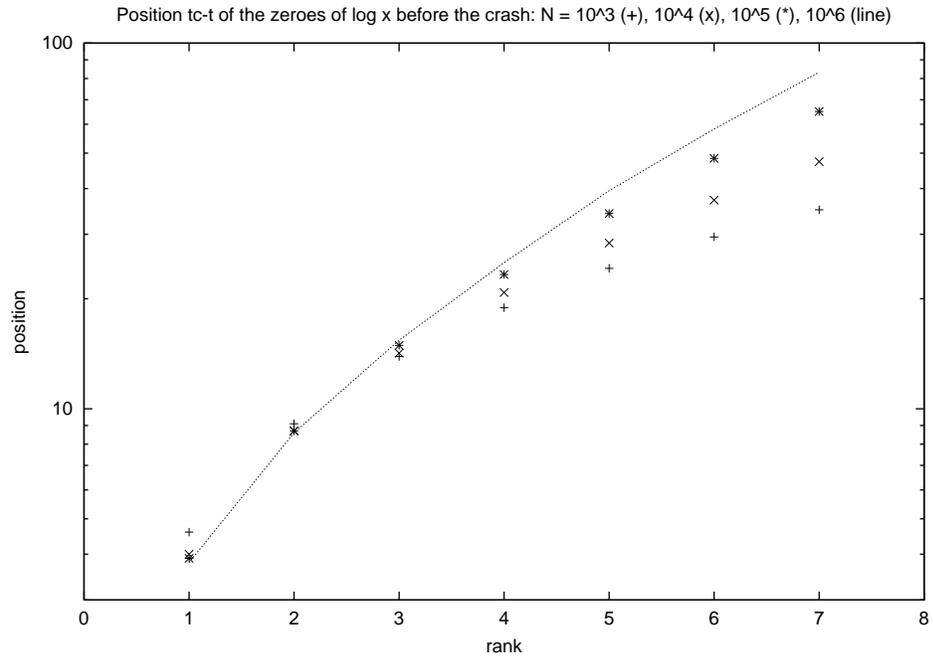}
\end{center}
\caption{
Position of zeroes in $x$ versus time shortly before the crash;
perfect log-periodic oscillations would correspond to straight lines in this
semilogarithmic plot and are achieved better for large markets. 
}
\end{figure}    

Qualitatively similar results are also observed with a Cont-Bouchaud type
of cluster trading model, if the amount of stocks to buy is proportional
to $x$, the solution of eq(6) with a small noise $\pm 0.002$ added to $x$. 

\section{Conclusion}

Most of the desired properties are recovered, if we combine in the Cont-Bouchaud
percolation model several of the modifications introduced in the past to get
specific properties. Thus this modified model has become quite realistic in
giving fat tails with proper exponent, multifractality, high-low asymmetry, 
volatility clustering and volatility-activity correlation. The crossover 
towards Gaussian behaviour needs some new ideas. In a different form it gives 
log-periodic oscillations before a crash.

Some of the simulations were made on the Cray-T3E of the Julich supercomputer
center. DS thanks the two other authors for their hospitality, partly financed
through KOSEF-DFG, during the time of this work.

\end{document}